\documentclass[utf8]{FrontiersinHarvard} 

\usepackage{url,hyperref,lineno,microtype,subcaption}
\usepackage[onehalfspacing]{setspace}
\usepackage{physics}
\usepackage{graphicx}
\usepackage{subcaption}
\usepackage{epstopdf}
\usepackage{soul}

\def\keyFont{\fontsize{8}{11}\helveticabold }
\def\firstAuthorLast{Speri {et~al.}} 
\def\Authors{Lorenzo Speri\,$^{1,*}$, Michael L. Katz$^{1,2}$, Alvin J. K. Chua\,$^{3,4,*}$, Scott A. Hughes\,$^{5}$, Niels Warburton\,$^{6}$, Jonathan E. Thompson\,${^7}$, Christian E. A. Chapman-Bird\,$^{8}$, Jonathan R. Gair\,$^{1}$}
\def\Address{$^{1}$Max-Planck-Institut f\"ur Gravitationsphysik, Albert-Einstein-Institut,
Am M\"uhlenberg 1, 14476 Potsdam-Golm, Germany \\
$^{2}$NASA Marshall Space Flight Center, Huntsville, Alabama 35811, USA\\
$^{3}$Department of Physics, National University of Singapore, Singapore 117551\\
$^{4}$Department of Mathematics, National University of Singapore, Singapore 119076\\
$^{5}$Department of Physics and MIT Kavli Institute, Cambridge, MA 02139, United States\\
$^{6}$School of Mathematics and Statistics, University College Dublin, Belfield, Dublin 4, Ireland\\
$^{7}$Theoretical Astrophysics Group, California Institute of Technology, Pasadena, CA 91125, U.S.A.\\
$^{8}$SUPA, University of Glasgow, Glasgow G12 8QQ, United Kingdom
}
\def\corrAuthor{Lorenzo Speri}
\def\corrEmail{lorenzo.speri@aei.mpg.de}

\def\corrAuthor{Lorenzo Speri, lorenzo.speri@aei.mpg.de}
\def\corrEmail{Alvin J. K. Chua, alvincjk@nus.edu.sg}

\begin{document}
\onecolumn
\firstpage{1}
\title {
Fast and Fourier: Extreme Mass Ratio Inspiral Waveforms in the Frequency Domain
} 

\author[\firstAuthorLast ]{\Authors} 
\address{} 
\correspondance{} 

\extraAuth{}

\maketitle

\begin{abstract}

\section{}
Extreme Mass Ratio Inspirals (EMRIs) are one of the key sources for future space-based gravitational wave interferometers.  Measurements of EMRI gravitational waves are expected to determine the characteristics of their sources with sub-percent precision. However, their waveform generation is challenging due to the long duration of the signal and the high harmonic content. Here, we present the first ready-to-use Schwarzschild eccentric EMRI waveform implementation in the frequency domain for use with either graphics processing units (GPUs) or central processing units (CPUs). We present the overall waveform implementation and test the accuracy and performance of the frequency domain waveforms against the time domain implementation. On GPUs, the frequency domain waveform takes in median $0.044$ seconds to generate and is twice as fast to compute as its time domain counterpart when considering massive black hole masses $\geq 2 \times 10^6 \,{\rm M_\odot}$ and initial eccentricities $e_0 > 0.2$. On CPUs, the median waveform evaluation time is $5$ seconds, and it is five times faster in the frequency domain than in the time domain. Using a sparser frequency array can further speed up the waveform generation, reaching up to $ 0.3$ seconds. This enables us to perform, for the first time, EMRI parameter inference with fully relativistic waveforms on CPUs.
Future EMRI models which encompass wider source characteristics (particularly black hole spin and generic orbit geometries) will require significantly more harmonics. Frequency-domain models will be essential analysis tools for these astrophysically realistic and important signals. 

\tiny
 \keyFont{ \section{Keywords:} gravitational wave, LISA, EMRI, data analysis, frequency domain} 
\end{abstract}

\section{Introduction}

The future space-based Laser Interferometer Space Antenna (LISA) will observe gravitational waves in the millihertz regime \citep{LISAMissionProposal,Gair:2022knq}. This region of the gravitational wave spectrum is rich in sources, including Galactic double white dwarf (WD) binaries \citep{Seto:2001pg}, massive black hole (MBH) binaries \citep{Klein_2015}, and stellar origin black hole (SOBH) binaries early in their evolution \citep{Klein_2015,Moore:2019pke}. Another key source class is extreme mass-ratio inspirals (EMRIs) \citep{Babak2007}. These are compact binaries with a mass ratio $\mu/M\simeq 10^{-4}-10^{-6}$, where $\mu\sim1-100M_\odot$ is the mass of the orbiting secondary and $M\sim10^{5}-10^{7}M_\odot$ is the mass of the MBH. EMRIs are expected to form in dense stellar clusters of galactic nuclei \citep{AmaroSeoane:2012tx, Pan2021EMRIFormation}, where their detection rate ranges from $\sim 1-10^4$ per year with observable signal-to-noise ratios (SNR) expected to be $\sim 20-1000$ over the duration of the signal \citep{Babak2017, Amaro-Seoane2007, Porter2009LISADataOverview}. The details depend on the precise formation mechanism \citep{Amaro-Seoane:2012lgq,Berry:2019wgg,Arca-Sedda2019, Gair2004EMRIsPreds}, but it is anticipated that the majority of EMRIs are expected to have eccentricities at plunge in the range $0 < e_p \lesssim 0.2$, with a long tail extending to larger $e_p$ \citep{Babak2017}. This high eccentricity is why EMRIs have some of the richest and most complicated gravitational waveforms of any compact binary system. The small mass ratio of EMRIs also means that they evolve slowly, typically completing $\sim 10^4-10^5$ orbits over years whilst in the LISA band. The long-lasting, complex waveforms of EMRIs present a substantial challenge for both the modeling of these binaries and the LISA data analysis task \citep{Babak_2009,Cornish_2011,MockLISADataChallengeTaskForce:2009wir,MockLISADataChallengeTaskForce:2007iof,MockLISADataChallengeTaskForce:2006sgi,Barack2018,Chua2022,Speri:2022kaq}.

Modeling and extracting EMRI signals from the LISA data stream will provide binary system measurements with sub-percent level precision \citep{Babak2017, Berry2019}. This precision will enable precise tests of general relativity \citep{Barack2007, Gair2013,Maselli:2020zgv,Maselli:2021men,Barsanti:2022ana}. Additionally, studying EMRIs will enhance our understanding of the mass function of MBHs \citep{Gair2010}, the dense stellar environments in galactic cores \citep{Amaro-Seoane2007}, and the gas disks surrounding MBHs \citep{Barausse2007, Barausse2008, Gair2011, Yunes2011, Barausse2014, Barausse2015, Speri:2022upm}. Furthermore, gravitational wave signals from EMRIs could be used to constrain cosmological parameters \citep{MacLeod:2007jd, Laghi2021EMRICosmology} or measure phase calibration errors \citep{Savalle:2022xpv}.

Extracting this wealth of information from EMRIs in the LISA data stream poses two significant challenges: accuracy and speed. The former is limited by the waveform modeling accuracy, whereas the latter is limited by the combined computational cost of the waveform generation, the discrete Fourier transform (DFT) and the inner product necessary to compute the likelihood function or any detection statistic.
A first attempt to address these problems was presented in the release of the \textsc{FastEMRIWaveforms} ({FEW}) computational framework \citep{Chua2020RapidGenLetter, Katz:2021yft}, which can compute fully relativistic waveforms rapidly. This package combines a set of standalone modules to create EMRI waveforms on both graphics processing unit (GPU) and central processing unit (CPU) hardware. It was shown that the GPU version is $\approx 2500$ times faster than its CPU counterpart for Schwarzschild spacetimes. GPUs are designed with a large number of cores and specialized hardware that allows them to perform multiple tasks simultaneously, making them highly efficient in handling parallel processing tasks. Compared to CPUs, which typically have fewer cores, GPUs can process a much larger number of tasks concurrently, resulting in significantly improved performance for certain applications. GPUs are particularly well-suited to efficient EMRI waveform generation because these signals consist of a large number of harmonic modes that are summed in parallel.

The waveform generation in {FEW} is currently performed in the time domain (TD), with typical speeds using GPUs and CPUs on the order of tens of milliseconds and tens of seconds respectively. For GPUs, the computational costs of the likelihood are negligible (less than a millisecond), therefore making the waveform generation the bottleneck of the computations. However, for CPUs all of the aforementioned operations contribute significantly to the overall data analysis cost. We expect that a frequency domain implementation of EMRI waveforms would reduce the likelihood evaluation cost. However, it is not immediately clear how the frequency domain implementation of the EMRI waveforms would comparatively perform on GPUs and CPUs.

In this work, we expand upon the {FEW} package by implementing a frequency domain (FD) EMRI waveform based upon~\cite{Hughes:2021exa}. We provide a module that can compute EMRI waveforms for eccentric inspirals into a non-rotating black hole at adiabatic order \citep{Hinderer2008twotimescale,Hughes:2021exa}. 
The implementation supports use with both CPUs and GPUs and its accuracy and performance are explored. This work is fully reproducible using the software provided at \url{https://github.com/lorenzsp/EMRI_FrequencyDomainWaveforms}

The paper is organised as follows. We review the general construction of EMRI waveforms in Section~\ref{sec:waveforms} and we discuss the frequency domain implementation in Section~\ref{sec:fourier}. In Section~\ref{sec:data_analysis} we present the data analysis tools that are used to test the accuracy of the waveform generation. Finally, in section \ref{sec:results} we present the results for the performance and accuracy tests of the frequency domain implementation.

\section{Methods}\label{sec:methods}

\subsection{Extreme Mass Ratio Inspiral Waveforms}\label{sec:waveforms}

The gravitational waves emitted by an EMRI are among the most complicated compact binary signals due to their long duration and rich harmonic content.
This complexity stems from the dynamics of the secondary object inspiralling into the central MBH. Such orbits are characterized by three degrees of freedom associated with the dimensionless fundamental frequencies of Kerr geodesic orbits $\Omega_{r,\theta,\phi}$\citep{Fujita2009FundFreqs,Schmidt:2002qk}. These quantities are determined by the dimensionless spin magnitude of the MBH $a$, and the quasi-Keplerian orbital parameters of $p$ (semi-latus rectum or separation in units of $M$), $e$ (eccentricity), and $x_I \equiv \cos{I}$ (cosine of the angle $I$ which describes the orbit's inclination from the equatorial plane). At adiabatic order, for a given system with primary mass $M$, secondary mass $\mu$, initial orbital parameters $(p_0,e_0,x_{I0})$, and initial phases  $\Phi_{\varphi 0}$, $\Phi_{\theta 0}$, $\Phi_{r 0}$, the orbital evolution can be determined by solving the following system of ordinary differential equations:
\begin{align*}
    \frac{d}{dt}p &= \frac{\mu}{M} \, f_p(a, p, e, x_I) \, \\
    \frac{d}{dt}e &=  \frac{\mu}{M} \, f_e(a, p, e, x_I) \, \\
    \frac{d}{dt}x_I &=  \frac{\mu}{M} \, f_{x_I}(a, p, e, x_I) \, \\
    \frac{d}{dt}\Phi_{\varphi,\theta,r} &= \Omega_{\varphi, \theta, r}(a, p, e, x_I)/M \, .
\end{align*}
The orbital-element fluxes $f_{p,e,x_I}$ account for the gravitational wave emission and are interpolated from pre-computed grids. In the {FEW} package, the system of ordinary differential equations is solved using an {explicit embedded Runge-Kutta (7, 8) method \citep{PRINCE198167}} with adaptive step size {(see \url{gsl_odeiv2_step_rk8pd} in \cite{gnu_library})}, resulting in sparse orbital trajectories of $\sim 100$ data points in length. 

Once the trajectory is fully specified by $\{p(t),e(t),x_{I}(t),\Phi_{\varphi}(t),\Phi_{\theta}(t),\Phi_{r}(t)\}$, we compute the complex gravitational-wave amplitudes $A_{lmkn}( p(t),e(t),x_I(t) )$. Each amplitude mode is associated with the multipole $l$ of the radiation and the harmonic $\{m,n,k\}$ of the fundamental orbital frequencies, where $m$, $k$, and $n$ label the azimuthal, polar, and radial modes, respectively. In the {FEW} package, the amplitudes and phases are interpolated using a ``not-a-knot'' cubic spline to handle the large number of modes using a sparse number of points. {Cubic splines are the simplest polynomial basis spline suitable for our purposes because they provide an approximation of the second derivative of frequency with respect to time. Higher-order spline functions are more expensive and might overfit data as sparsely evaluated as our trajectory information. Additionally, the root-finding procedure that will be used in constructing the time-frequency correspondence in Eq.~\ref{eq:cubic_equation} is significantly more expensive for higher-order polynomials due to the need to perform this task numerically.}
In the interest of computational efficiency, we control the number of harmonic modes with a further mode-selection step. {The mode-selection step is defined by the threshold $\epsilon$ of the total power emitted by all modes}. {Within each time step of the sparse array, we take $|A_{lmkn}(t)|^2$ and sort the individual modes in descending order and compute a cumulative summation. From this sorted array, we check whether the additional power contributed by an individual mode falls below the threshold $\epsilon$ related to the total power emitted. This gives a set of contributing modes within each time step. The selected modes are given by the union of all contributing modes across all time steps. This ensures to maintain continuity across time.} Only the modes that pass this threshold are included in the waveform computation (see \cite{Katz:2021yft} for further details).

The gravitational wave signal at a large distance from the source can be represented in terms of the complex time-domain dimensionless strain \citep{Hughes:2021exa}:
\begin{equation}\label{eq:main_wave}
    h =h_+-ih_\times= \frac{\mu}{d_L}\sum_{lmkn} A_{lmkn}(t)S_{lmkn}(t, \theta)e^{im\phi} \exp\qty{-i\qty[ m \Phi_\varphi(t) + k \Phi_\theta(t) + n\Phi_r(t) ] },
\end{equation}
where $t$ is the time of arrival of the gravitational wave at the Solar System barycenter, $\theta$ is the source-frame polar viewing angle, $\phi$ is the source-frame azimuthal viewing angle, and $d_L$ is the  luminosity distance of the source.  The functions $S_{lmkn}(t,\theta)$ are spin-weighted spheroidal harmonics.  For inspiral into a Schwarzschild black hole (the focus of this analysis), these functions are identical to the spin-weighted spherical harmonics.  For the general Kerr case, they differ from the spherical harmonics (though they can be very usefully expanded into spherical harmonics \citep{Hughes2000}), and in particular they depend on the orbital frequencies.  Because the orbital frequencies evolve with time, the spheroidal harmonics evolve as well \citep{Hughes:2021exa}, complicating the description of the waveform.  Future work will describe how we handle this complication.  In any case, the usual transverse-traceless gravitational wave polarizations can be obtained from the real and imaginary parts of the complex conjugate of $h$.

A key element to be noted in Eq.~(\ref{eq:main_wave}) is that the EMRI waveform can be viewed as a sum of simple “voices”, where each voice corresponds to a mode $(l,m,k,n)$. The time-domain waveform can thus be rewritten in terms of a sum over voices $V$:
\begin{equation}\label{eq:wave_voice}
    h = \sum_{V} H_V(t) e^{-i\Phi_{V}(t)},
\end{equation}
where each voice is characterized by a complex amplitude $H_V(t)$ and a phase $\Phi_{V}(t)$. The voice-by-voice decomposition was suggested to one of the present authors by L. S. Finn, and first presented in \cite{Hughes:2001jr}\footnote{Note that a frequency domain waveform implementation was also presented in \cite{PhysRevD.107.024021} with the limitation to circular Kerr orbits.}. In this work, we provide the first ready-to-use implementation of this decomposition for frequency-domain EMRI waveforms.

\subsection{The frequency domain}
\label{sec:fourier}
EMRI waveforms can be represented in the frequency domain by using the stationary phase approximation. Because amplitude, phase, and frequency in an EMRI evolve slowly (as long as the two-timescale approximation is valid \citep{Miller:2020bft}), we expect the stationary phase approximation (SPA) to provide a high-quality approximation to the Fourier transform of the signal \citep{Droz:1999qx}. Here, we review the computation of the Fourier transform of an EMRI waveform in the frequency domain presented in~\cite{Hughes:2021exa}. We begin by taking the Fourier transform of each voice of Eq.~(\ref{eq:wave_voice}) as follows:
\begin{eqnarray}
{\tilde h}(f) &\equiv& \int_{-\infty}^\infty h(t)e^{2\pi i f t}\,dt
= \int_{-\infty}^\infty H(t) e^{i[2\pi f t - \Phi(t)]}\,dt\;
\label{eq:singlevoiceFT}
\end{eqnarray}
where we have not written the subscript for ease of notation. 
To compute the stationary phase approximation to the signal, it is necessary to expand the phase evolution, written as
\begin{equation}
\Phi(t) = \Phi(t_S) + 2\pi F (t - t_S) + \pi \dot F (t - t_S)^2 + \frac{\pi}{3} \ddot F (t - t_S)^3+\ldots\,
\label{eq:phaseexpand2}
\end{equation}
where
\begin{equation}
F \equiv \frac{1}{2\pi}\frac{d\Phi}{dt}\biggr|_{t_S}\;,\qquad
\dot F = \frac{dF}{dt} \equiv \frac{1}{2\pi}\frac{d^2\Phi}{dt^2}\biggr|_{t_S}\;,
\qquad
\ddot F \equiv \frac{1}{2\pi}\frac{d^3\Phi}{dt^3}\biggr|_{t_S}\;.
\end{equation}
We have introduced the signal's instantaneous frequency and the instantaneous first and second derivatives of the frequency at the stationary time $t = t_S$ for which $F = f$. 
We expanded up to third order in the phase evolution because for some voices the frequency evolution is not monotonic. If the frequency derivative vanishes ($\dot F=0$), the standard SPA Fourier transform is singular and cannot be used \citep{Klein:2014bua}. This is highly relevant to EMRI signals because the frequency associated with many voices rises to a maximum and then decreases. In particular, this occurs for EMRI voices which involve harmonics of the radial frequency. This is because $\Omega_r$ becomes very small (approaching $|\Omega_\phi|$) in the weak-field limit, and goes to zero as systems approach the last stable orbit; a maximum of $\Omega_r$ exists between these two ``small'' limits.

We now use Eq.~(\ref{eq:phaseexpand2}) to obtain the time-frequency correspondence for each voice, i.e. we find the time $t$ such that $F(t) = f$. The frequency evolution of each voice $F(t) =  (m \Omega_\varphi(t) + k \Omega_\theta(t) + n\Omega_r(t))/(2 \pi M ) $ is directly obtained from the trajectory evolution of each of the fundamental frequencies. During the waveform production process, the fundamental frequency evolution is interpolated with a cubic spline allowing for a good approximation using a third-order polynomial. Doing so, the condition $f=F(t)$ between two knots can be rewritten as:
\begin{equation}\label{eq:cubic_equation}
    f = A + B \, (t - t_i) + C \, (t - t_i) ^2 + D \, (t - t_i)^3 , \,
\end{equation}
where $A, B, C, D$ are real numbers given by the sum of the spline coefficients of three frequencies multiplied by their mode number, and $t_i$ is the time at the beginning of the spline segment. 

The fundamental frequency, fundamental phase, and amplitude spline coefficients are determined along the sparse trajectory prior to the final waveform summation phase (see \cite{Katz:2021yft} for more information). Additionally, using all values of $f$ and each set of frequencies found along the sparse trajectory for each voice, the spline segment in which each value of $f$ falls is predetermined before the summation. Once in the summation, the goal is to determine $t(f)$ as efficiently as possible. This is difficult in cases where the voice is not monotonic because $t(f)$ is not a bijective mapping for this case: two values of $t$ can be represented by the same value of $f$. To address this issue, we solve for the cubic roots of Eq.~\ref{eq:cubic_equation} within each segment where $f$ is found (one or two segments per voice depending on $f$). We take the real-valued roots and determine which one or two actually exist between the bounds of the segment we are currently examining (it can be two only for the segment in which the turnover occurs). Since the frequency evolution is slowly varying within each segment, the condition $\dot F = 0$ cannot occur more than one time per segment. These time values occurring within our segment are then used to determine all necessary quantities for waveform building because all informative splines are fit as a function of $t$.

We define $t_j(f)$ to be the $j$th time (first or second) at which $F(t) = f$, and write $\dot F_j \equiv \dot F[t_j(f)]$, $\ddot F_j \equiv \ddot F[t_j(f)]$. These higher-order derivatives are also computed using the fundamental frequency spline information. With that, for each value of $f$ for each voice, the Fourier transform can be represented as:
\begin{eqnarray}
{\tilde h}(f) &\simeq& \sum_{j=1}^NH[t_j(f)]e^{i[2\pi ft_j(f) - \Phi(t_j(f))]}\times \int_{-\infty}^\infty e^{-i \pi[\dot F_j(t')^2 + \ddot F_j(t')^3/3]}dt'\;,
\label{eq:SPAintegral2}
\end{eqnarray}
where $N\in\{1, 2\}$ is the number of time values associated with each value of $f$. To perform this integral, we set $\alpha = \gamma + 2\pi i \dot F$, with $\gamma$ real and positive, define $\beta = 2\pi \ddot F$, and use
\begin{equation}
\int_{-\infty}^\infty e^{-\alpha t^2/2 - i \beta t^3/6}\,dt = \frac{2}{\sqrt{3}}\frac{\alpha}{|\beta|}e^{\alpha^3/3\beta^2} K_{1/3}(\alpha^3/3\beta^2)\;,
\label{eq:Iintegral}
\end{equation}
where $K_n(z)$ is the modified Bessel function of the second kind. Taking the limit $\gamma \to 0$, we find

\begin{equation}
{\tilde h}(f) \simeq \frac{2}{\sqrt{3}}\sum_{j=1}^NH[t_j(f)]e^{i[2\pi ft_j(f) - \Phi(t_j(f))]}\frac{i\dot F_j}{|\ddot F_j|}e^{-2\pi i\dot F_j^3/3\ddot F_j^2}K_{1/3}(-2\pi i \dot F_j^3/3\ddot F_j^2)\;.
\label{eq:extendedspa1}
\end{equation}
This result defines our frequency domain waveform.  Expanding about ``small'' $\ddot F_J$, it can be shown that this yields the standard SPA when $\ddot F_j \rightarrow 0$ and $N = 1$ (see \cite{Hughes:2021exa}). {For computational efficiency, we approximate the term $i \sqrt{x} e^{-i X} K_{1/3}(-i X)$ with two polynomial expansions in $X = 2\pi i \dot F_j^3/3\ddot F_j^2$, valid for $|X|<7$ and $|X|>7$. The largest relative error occurs around the transition region $|X|\approx 7$ and is $\approx 10^{-3}$.}
 {We did not explore the potential of applying this method for calculating other oscillatory integrals \citep{fcfio_Candes2007}.}

As previously mentioned, this waveform is built into the framework of {FEW}. All parts of the waveform generation prior to the ``waveform summation module'' are identical to the original time-domain implementation \citep{Katz:2021yft}. The waveform summation described above is implemented for both CPUs and GPUs. For the purposes of GPU parallelization, the summation is grouped by harmonic voice and segment of the splines defined in time. This creates a rectangular computation grid because all spline information occurs on the same sparse time evolution returned from the trajectory integrator yielding a constant number of segments across all spline constituents. On the GPU, we place each segment within each voice on a separate GPU block (larger parallelization unit). Within each block, we use the parallel threads (smaller parallelization unit) to evaluate all values of interest in frequency for that specific voice and segment. The waveform is then constructed by ``atomically'' adding from each of these blocks and threads to an overall waveform vector stored in GPU global memory.


\subsection{Data analysis setup}\label{sec:data_analysis}
An EMRI waveform for non-rotating black holes is uniquely determined in the reference frame of the Solar System barycenter by setting the following parameters:
\begin{equation}
    \{M, \mu, p_0 , e_0, d_L, \theta_S, \phi_S, \theta_K, \phi_K, \Phi_{\varphi 0}, \Phi_{r 0}\} \, ,
\end{equation}
where $\theta_S$ and $\phi_S$ are the polar and azimuthal sky location angles given in the Solar System barycenter reference frame, and $\theta_K$ and $\phi_K$ are the azimuthal and polar angles describing the orientation of the orbital angular momentum. The relation to the source frame angles $\theta,\phi$ can be found in \cite{Katz:2021yft}. We do not include the effect of the detector response function, but we leave this to future works \citep{Marsat:2018oam}.

We check the performance and accuracy of the FD waveform over the EMRI parameter space confined to $\log (M/{\rm M}_{\odot})\in [\log 10^{5},\log  10^{7})$, $\log (\mu/M) \in [\log 10^{-6},\log 10^{-4})$ , $e_0\in [0.001, 0.7)$ and $\Phi_{\varphi 0}, \Phi_{r 0} \in [0, 2\pi)$ by uniformly drawing 5000 parameter realizations. The initial semi-latus rectum is fixed to obtain an inspiral with a time duration of $0.99T$, where $T$ is the duration of the observation.
We consider $h_+$ and $h_\times$ in the Solar System barycenter as our two channels, and we fix $\{d_L=1\,{\rm Gpc}, \theta_S = \pi/3, \phi_S= \pi/3, \theta_K= \pi/3, \phi_K= \pi/3 \}$ throughout this work.
To check our implementation, we quantify how similar two waveforms are using the mismatch\footnote{In the literature this is also denoted unfaithfulness.} 
\begin{equation}\label{eq:mism}
\mathcal{M} (a,b) = 1-\sum_{\alpha} \frac{\bra{a}\ket{b}_{\alpha}}{\sqrt{\bra{a}\ket{a}_{\alpha}\bra{b}\ket{b}_{\alpha}}} \, 
\end{equation}
where we have introduced the inner product $\bra{\cdot}\ket{\cdot}$ of each channel $\alpha$,
\begin{equation}\label{eq:innerprod}
\bra{a (t)}\ket{b (t)} =4 \Re \int _{0} ^\infty \frac{\tilde{a} ^* (f) \tilde{b} (f) }{S_n (f)} \, \dd f \approx 4 \Re \sum _{i} \, \frac{\tilde{a} ^* (f_i) \tilde{b} (f_i) }{S_n (f_i)} \, \frac{1}{T} \, ,
\end{equation}
where $f_i = i/T$, with $i$ from 1 up to the total {number of frequency bins} $N_f = N/2$ with $N$ the number of time points. Note that we will also consider frequency arrays with spacing different from the one previously mentioned.
The tilde indicates the Fourier transform, and $S_n (f)$  is the one-sided noise spectral density for the LISA data streams, which we take from \cite{Babak:2021mhe}. We use the mismatch $\mathcal{M} (h_{\rm FD},h_{\rm TD})$ to compare the frequency domain implementation with the DFT of the time-domain implementation sampled with a time interval $\Delta t = T/N$. 
{The conversion from continuous to discrete samples changes the underlying Fourier transform into the discrete Fourier transform. This leads to a type of distortion called aliasing. The choice of an appropriate time sampling $\Delta t$ is the key to minimizing such distortion. The Nyquist rate is the maximum resolvable frequency for the DFT of a discrete-time signal and its value equals $1/(2\Delta t)$. When the signal is digitized, we need to make sure that the Nyquist frequency is larger than the signal's largest harmonic frequency $F_V(t)$, i.e. $\text{max}_{V} F(t)<1/(2\Delta t)$, where $V=(l,m,n,k)$. By ensuring this condition, the resulting discrete-time sequence is free of aliasing.
Similarly, the conversion from a very long (or infinite) sequence to a manageable size entails a type of distortion called leakage, which is manifested as a loss of  resolution in the DFT.} We mitigate this effect by applying the Hann window to the time domain waveform \citep{2020SciPy-NMeth}. To account for the effect of windowing in the FD waveform, we perform a convolution of the FD signal with the DFT of the window function. We do not maximize over the phase since the initial conditions of the frequency and phase evolution of the TD and FD waveforms are the same by construction. Global relative time and phase shifts are not observable parameters, so by not optimizing over them we are being conservative with the model and folding data processing into 
the model comparison.

\begin{figure}

\caption{Spectrum of the frequency domain (FD) waveform compared to the spectrum of the discrete Fourier transform (DFT) of the time domain (TD) waveform. In the upper plot, we show the amplitude squared $|\tilde h_{+} (f)|^2$ when we do not include any windowing when computing the spectrum of the waveforms. In the lower panel, we apply the Hann window to both the TD and FD waveforms.
{Note how the windowing helps to resolve the harmonics in the low-frequency region $f<10^{-4}$ Hz and reduce the power of the leakage at frequencies $f>10^{-2}$ Hz.}
We consider an EMRI system composed of a secondary object with mass $\mu=10 {\rm M}_{\odot}$ orbiting around a non-rotating black hole of mass $M=10^6 {\rm M}_{\odot}$ with initial eccentricity $e_0=0.6$ and semi-latus rectum $p_0=9.56$ fixed such that the inspiral plunges after 1 year of observation.
}

  \includegraphics[width=0.9\linewidth]{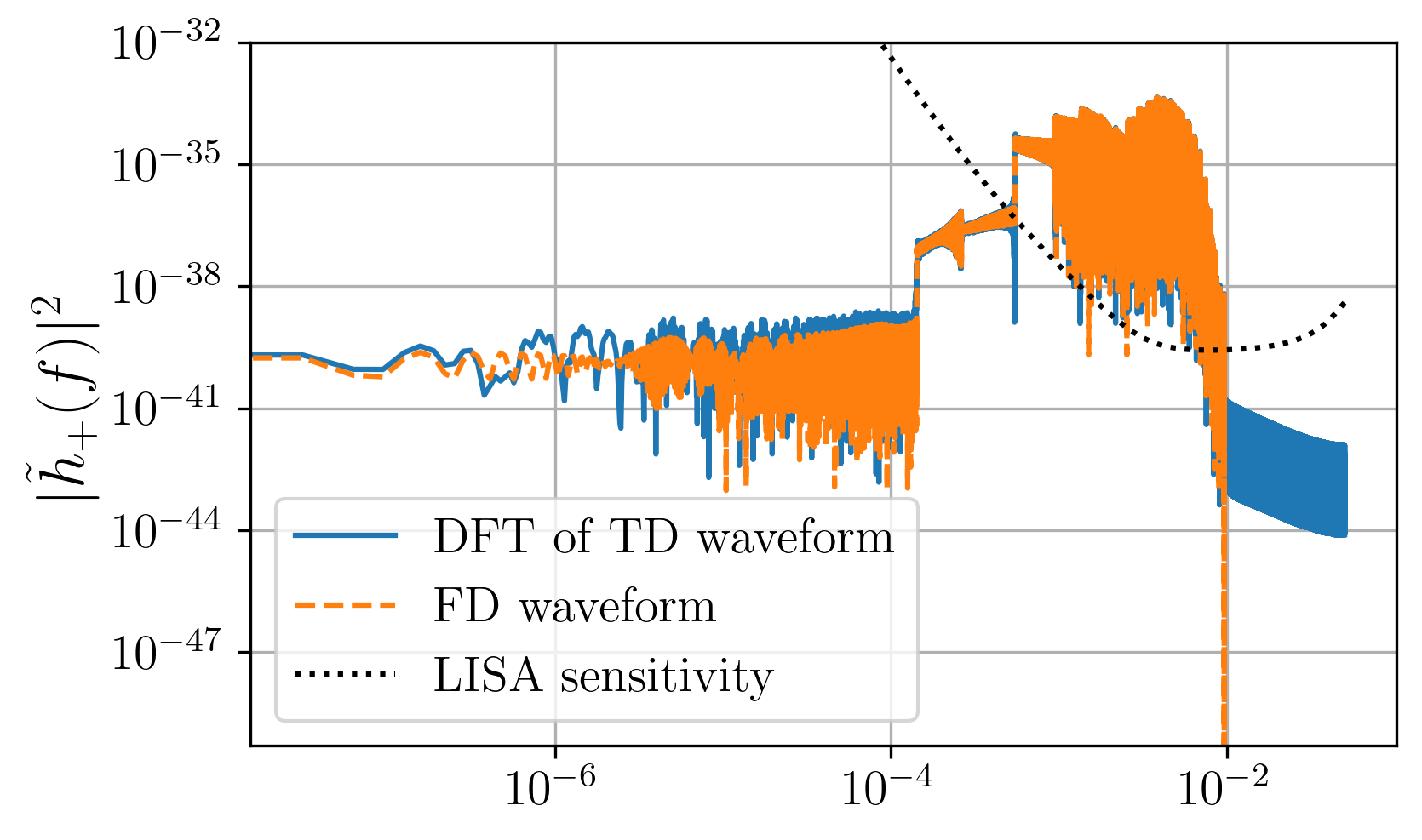}
  \includegraphics[width=0.9\linewidth]{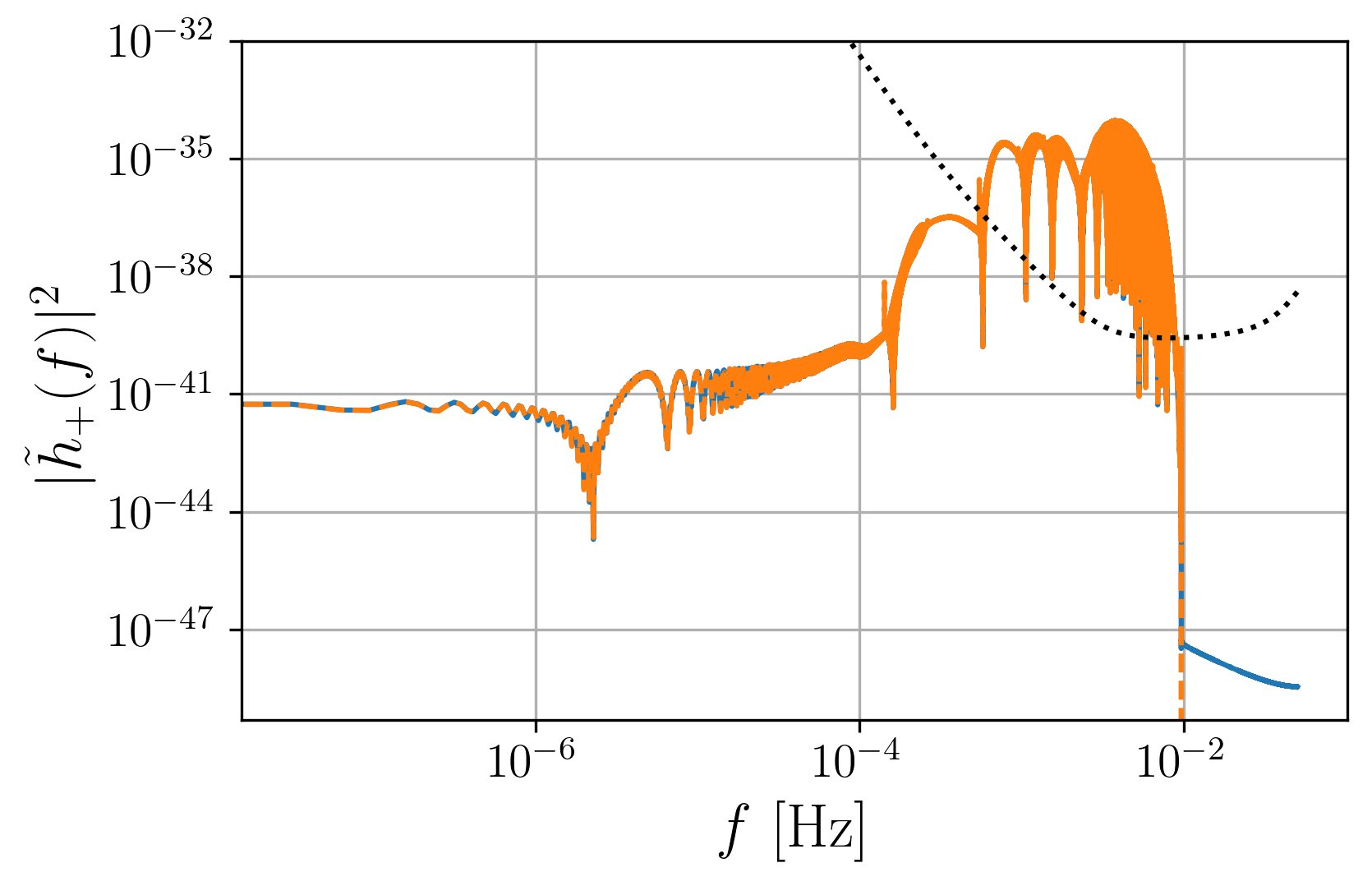}
\label{fig:waveforms}
\end{figure}

We show the impact of windowing and spectral leakage in Figure~\ref{fig:waveforms} for a system with $\mu=10 {\rm M}_{\odot}$, $M=10^6 {\rm M}_{\odot}$, $e_0=0.6$, $p_0=9.56$ for an observation of $T=1$ year, a mode content threshold $\epsilon=10^{-2}$, and a sampling interval of $\Delta t=10$ seconds. 
We show the amplitude squared, $|\tilde{h}_{+}(f)|^2$, of the frequency domain waveform (dashed orange line) and of the DFT of the time domain waveform (solid blue line) for the plus polarization in the upper panel of Figure~\ref{fig:waveforms}. For reference, we also show the LISA sensitivity curve used to calculate the inner products. 
In the upper panel of Figure~\ref{fig:waveforms} we do not apply any windowing and therefore the FD waveform is zero only where there is no mode content. Conversely, the DFT of the TD waveform is affected by spectral leakage, especially for frequencies larger than $10^{-2}\,$Hz. If we apply the Hann window consistently to both the FD and TD signals as previously described, we obtain the spectrum shown in the lower panel of Figure~\ref{fig:waveforms}. The effect of windowing reduces the total power but improves the agreement between the waveforms such that the mismatch decreases from $8\times 10^{-4}$ to $4\times 10^{-6}$.

It is often quoted that the average mismatch for a waveform needs to be $1-0.9^{1/3} \approx 1-0.965$ to recover 90\% of the signals in an experiment \citep{Owen:1998dk,Apostolatos:1995pj}. However, the requirements on the mismatch for parameter estimation (PE) studies are not as straightforward to characterize, primarily as there is no one-to-one relationship between the level of mismatch and the level of bias that is incurred during inference. A sensible requirement for a PE study is to ask that the systematic (or mismodeling) error coming from an approximate waveform is lower than the statistical error coming from the noise in the data. While the former is independent of the signal-to-noise ratio ($\text{SNR} = \sqrt{\bra{h_{+}}\ket{h_{+}} + \bra{h_{\times}}\ket{h_{\times}} }$), the latter does depend on the SNR, and therefore the mismatch requirement on a waveform for a PE study is SNR-dependent. {The mismatch times the SNR$^2$ is approximately half the model waveform error: $\mathcal{M}(h_1,h_2) \times {\rm SNR}^2 \approx \bra{h_1-h_2}\ket{h_1-h_2}/2$ \citep{Lindblom:2008cm}.}
Therefore, we perform PE on the system that yields the largest value of $\mathcal{M} \times {\rm SNR}^2$. 

In particular, we check that the posterior distribution obtained with an FD waveform is equivalent to the one obtained with a TD waveform. To do this, we use the likelihood given by \citep{whittleAnalysisMultipleStationary1953}:
\begin{equation}
    \label{eq:likelihood}
    p(s|h) \propto \exponential \qty [-\frac{1}{2} \sum_{\alpha = +,\times} \bra{s - h}\ket{s- h}_\alpha ]\, ,
\end{equation}
where we ignored the normalization factor that needs to be included when fitting for the shape of the noise spectral density $S_n$.
We do not inject any noise in the datastream since we want to check whether any waveform systematic arises due to the difference between the FD and TD implementations. Note that if we apply a window to our data, we should take into account its effects in the likelihood. In fact, the expectation value of the inner product of the noise $n(t) = s(t) - h(t)$ is different if a window is applied
\begin{equation}
\mathbb{E}\qty[\bra{w(t)\, n(t)}\ket{w(t)\,n(t)}] \neq
\mathbb{E}\qty[\bra{n(t)}\ket{n(t)}] = N
\end{equation}
We do not take into account the effect of windowing in the likelihood because this would lead to a non-diagonal noise covariance matrix and an increase in the computational cost of the inner product.
This is not a problem for the purpose of our study because we are consistently neglecting the effect of windowing in the likelihood when analyzing the data with different waveform models (FD or TD) \citep{BurkeTh}.
We remind the reader that we apply the windowing solely because the two waveforms (FD and TD) are in two different domains and we perform the parameter estimation only to check the presence of waveform systematics. 
However, we remark that the effect of windowing (and gaps) in LISA data is an important issue that remains to be addressed, but it is beyond the scope of this work. {To deal with the data gaps expected in LISA it would be worth investigating the usage of nonuniform DFT.}

\section{Results}\label{sec:results}
The frequency and time domain waveforms are built using the same trajectory and amplitude modules. However, the final signal output is obtained using different mode summations and prescriptions. Therefore, any waveform difference might be attributed to the stationary phase approximation and/or spectral leakage. However, any difference in the speed of the waveform generation is due to the mode summation algorithm of the two domains. In the following, we investigate the waveform mismatch and speed by comparing the time and frequency domain implementations over the parameter space for different observation time-spans $T$, sampling intervals $\Delta t$ and mode content thresholds $\epsilon$.
We consider as our fiducial values an observation time-span of $T=4$ years (yrs), a sampling interval of $\Delta t = 5$ seconds, and a threshold $\epsilon=10^{-5}$ and we vary each of these one at a time.

\subsection{Accuracy  analysis}\label{sec:mismatch}
\begin{figure}[]
\caption{ \emph{Upper panel}: Mismatch between the frequency domain (FD) and time domain (TD) waveforms over the parameter space of EMRI systems and for different observation time-spans, $T$, sampling intervals, $\Delta t$ and mode content thresholds, $\epsilon$. The different parameter realizations are drawn uniformly from the following ranges: $\log (M/{\rm M}_{\odot})\in [\log 10^{5},\log  10^{7})$, $\log (\mu/M) \in [\log 10^{-6},\log 10^{-4})$ , $e_0\in [0.001, 0.7)$, $\Phi_{\varphi 0}, \Phi_{r 0} \in [0, 2\pi)$. \emph{Lower panel}: Dependence of the mismatch on the central black hole mass $M$. The points that abruptly rise to larger mismatches for the orange dots are caused by the fact that, for some low-mass systems, the sampling interval $\Delta t =10$ seconds is not small enough to resolve the largest frequency.
}

\centering
\includegraphics[width=0.95\linewidth]{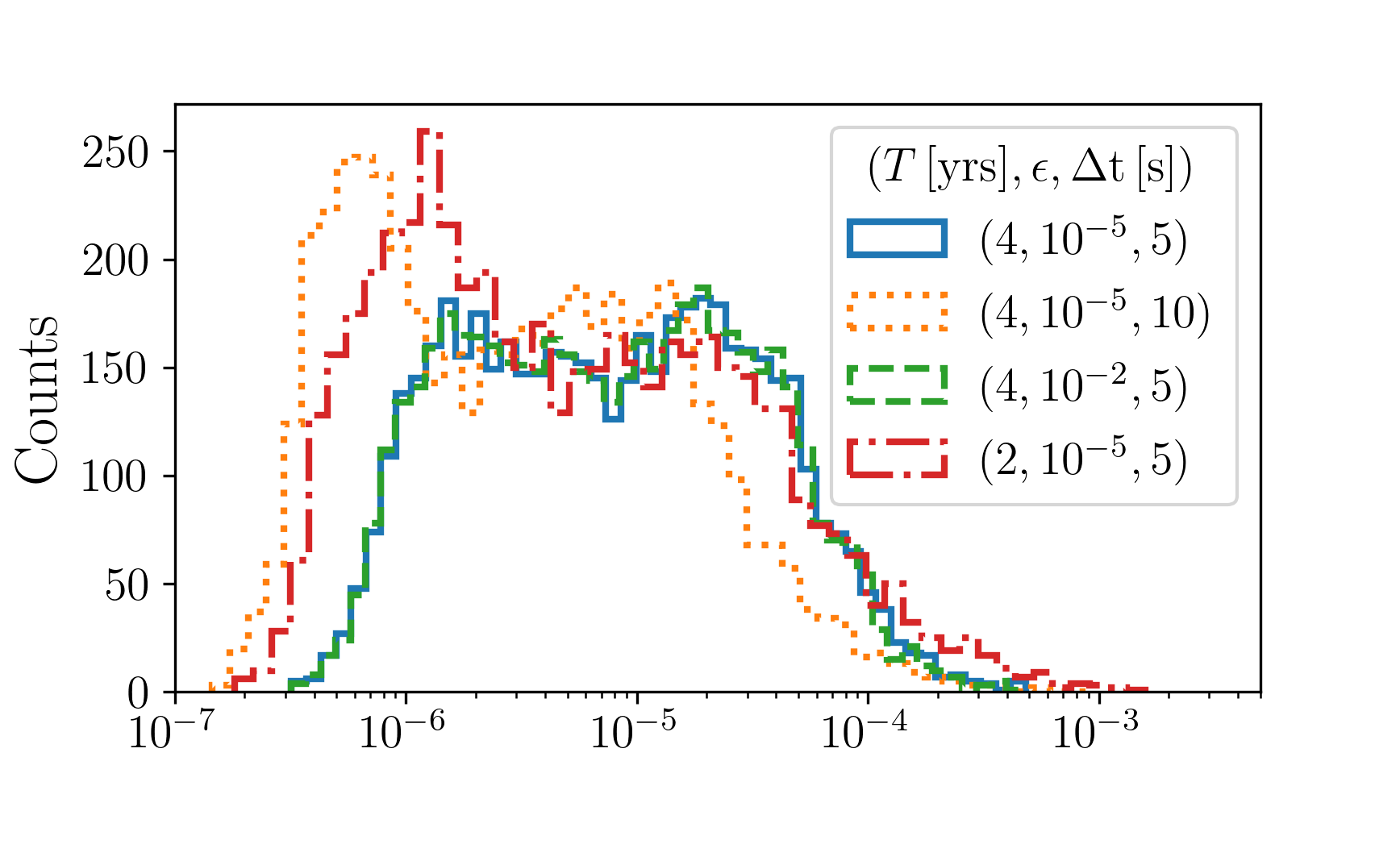}
\includegraphics[width=0.95\linewidth]{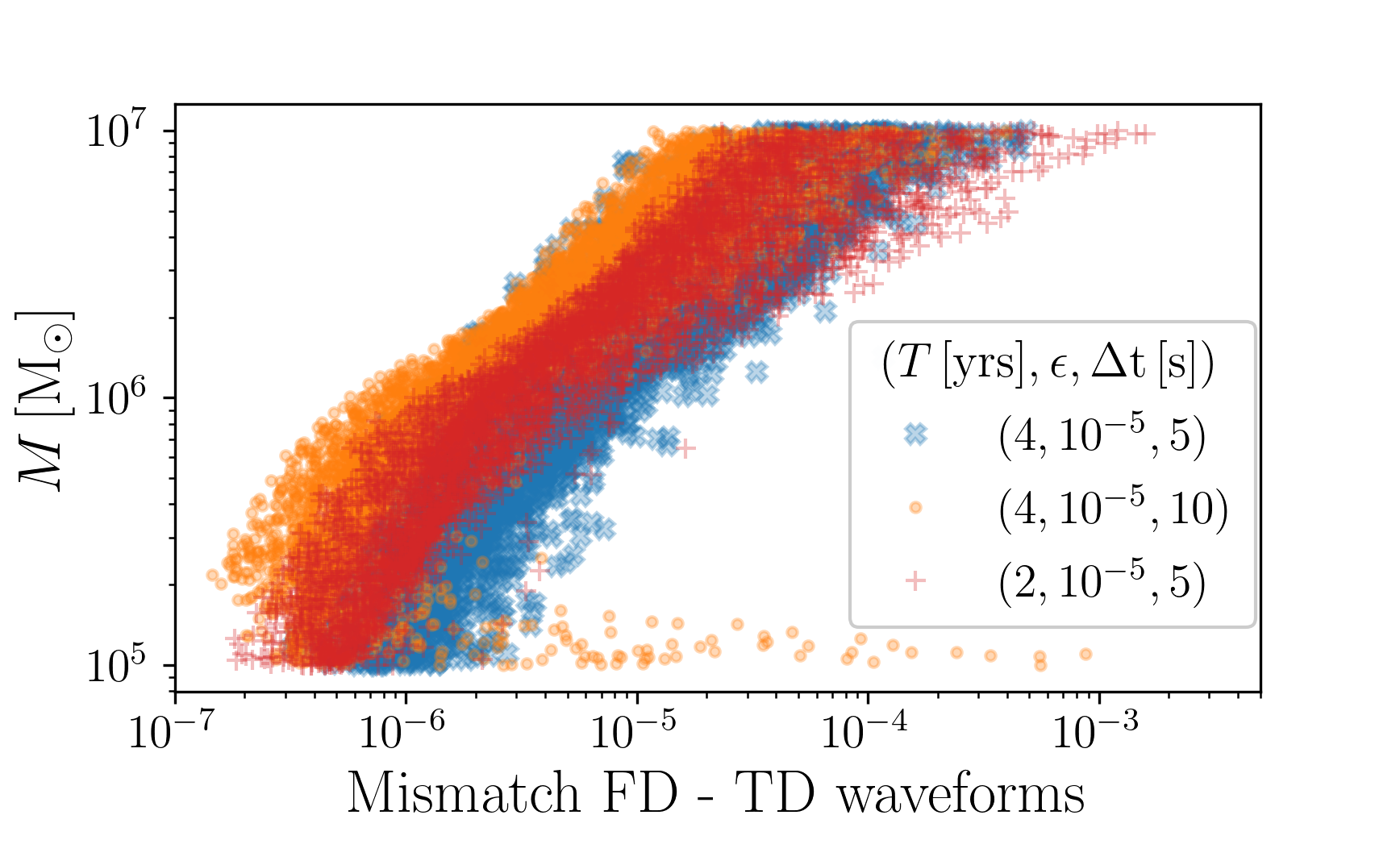}
\label{fig:mismatch}
\end{figure}

We calculate the mismatch between the frequency and time domain waveforms as described in Sec.~\ref{sec:data_analysis}. The mismatches obtained over the parameter space are shown in the upper part of Figure~\ref{fig:mismatch}.
The median of the mismatch distributions are between $4\times10^{-6}$ and $10^{-5}$, whereas the largest 95\% quantile of the mismatch is $9.7 \times10^{-5}$. 

The distributions obtained with a different mode content $\epsilon=10^{-5}$ (solid blue histogram) and $\epsilon=10^{-2}$ (dashed green histogram) are almost identical. This is expected since these systems are affected in the same way by spectral leakage.
When comparing the mismatches obtained with different sampling intervals $\Delta t=5$ s (solid blue histogram) and $\Delta t=10$ s (dotted orange histogram) we obtain lower mismatches for larger sampling intervals. We attribute this difference to the fact that the total number of {points} $T/\Delta t$ contributing to the mismatch is different. For the systems with lower $T/\Delta t$, the spectral leakage is weaker since the total power is lower. To confirm this we take the ratio of the mismatches $\mathcal{M}_{\Delta t=10 {\rm s}}/\mathcal{M}_{\Delta t=5 {\rm s}} $ and we find that it to be approximately constant as a function of the mass. For the same reason the mismatches obtained for $T=2$ years (dashed-dotted red histogram) are slightly lower than the ones obtain for the fiducial value $T=4$ years (solid blue histogram). However, the mismatches obtained for $(T,\Delta t)=(2\, {\rm yrs}, 5\, {\rm s})$ (dashed-dotted red histogram) are slightly larger than the ones obtained for $(T,\Delta t)=(4 \, {\rm yrs}, 10 \,  {\rm s})$(dotted orange histogram). This is probably due to the fact that we fix $p_0$ for a given inspiral duration, and the maximum $p_0$ allowed by the current implementation is $p_0\approx 16$.

In the lower panel of Figure~\ref{fig:mismatch} we show the behavior of the mismatch as a function of the central black hole mass $M$. The correlation coefficients between the mass $M$ and the mismatch $\mathcal{M}$ is approximately $\sim 0.6$. This is caused by the shape of the LISA sensitivity curve and the fact that for larger $M$ systems the spectrum shifts to lower frequencies and the spectral leakage present at the high frequencies becomes more dominant. For large $M$ the fiducial distribution of mismatches (blue crosses) is wider than the one obtained with $\Delta t= 10$ seconds, but both distributions reach approximately the same mismatches of $\sim 10^{-5}$ for $M\sim 10^7$. This is no longer the case when $M$ becomes smaller. For $M \sim 10^{5}$, the mismatches of $\Delta t =10$ seconds are generally smaller than the ones obtained from $\Delta t = 5$ seconds\footnote{The points that abruptly rise to larger mismatches for the orange distribution are caused by the fact that, for some low mass systems, the sampling interval $\Delta t =10$ seconds is not small enough to resolve the largest frequency.}. This confirms that for larger masses the spectral leakage dominates, whereas for smaller masses the mismatch is more affected by the total power in the signal. This behavior is also confirmed for systems with $T=2$ years (red crosses). In fact, if we compare the systems with $(T,\Delta t) =(4 \,{\rm yrs},10\, {\rm s})$ and $(T,\Delta t) =(2\, {\rm yrs},5\, {\rm s})$ that have the same power, the  distribution for the latter system reaches larger values than the distribution for the former system at all masses because the maximum resolvable frequency, $1/(2\Delta t)$, is higher. We note that for lower masses, the  distribution for $(T,\Delta t) =(2\, {\rm yrs},5\, {\rm s})$ has lower mismatches than the distribution for $(T,\Delta t) =(4\, {\rm yrs},5\, {\rm s})$ because the total power $T/\Delta t$ is smaller.

\begin{table}[htbp]
\caption{EMRI parameters of the source with largest $\mathcal{M} \times {\rm SNR}^2$, where $\mathcal{M}$ is the mismatch between the frequency and time domain waveforms. The posteriors distributions obtained with different waveform implementations are shown in Fig.~\ref{fig:posterior_td_fd} for a mode content of $\epsilon = 10^{-5}$.}
\centering
\begin{tabular}{c|c|c|c|c|c}
$M \,[ M_\odot]$ & $\mu \,[ M_\odot]$ & $p_0 \,[M]$ & $e_0$ & $T $[yrs] &SNR \\[0.1cm]
\hline
$3.67004174 \times 10^{6}$ & $292.058317$ & $13.7091019$ & $0.579413083$ & 4 &$78$ \\
\end{tabular}

\label{tab:source}
\end{table}

\begin{figure}[h]
\caption{Posterior distributions obtained with different waveform implementations for an EMRI system with parameters given in Table~\ref{tab:source}. In red we show the posterior distribution obtained using a frequency domain (FD) template to recover an injected waveform generated in the time domain (TD). The red distribution matches the posteriors obtained when the injected and template waveforms are in the same domain (blue and green posteriors).
The observation time, the sampling interval and mode content are $(T, \Delta t, \epsilon)=(4 \,{\rm yrs}, 10 \,{\rm s}, 10^{-5})$. The Kullback-Leibler divergences between each pair of marginal posteriors are of order $10^{-4}$ for all the considered parameters.
}

\centering
\includegraphics[width=0.95\linewidth]{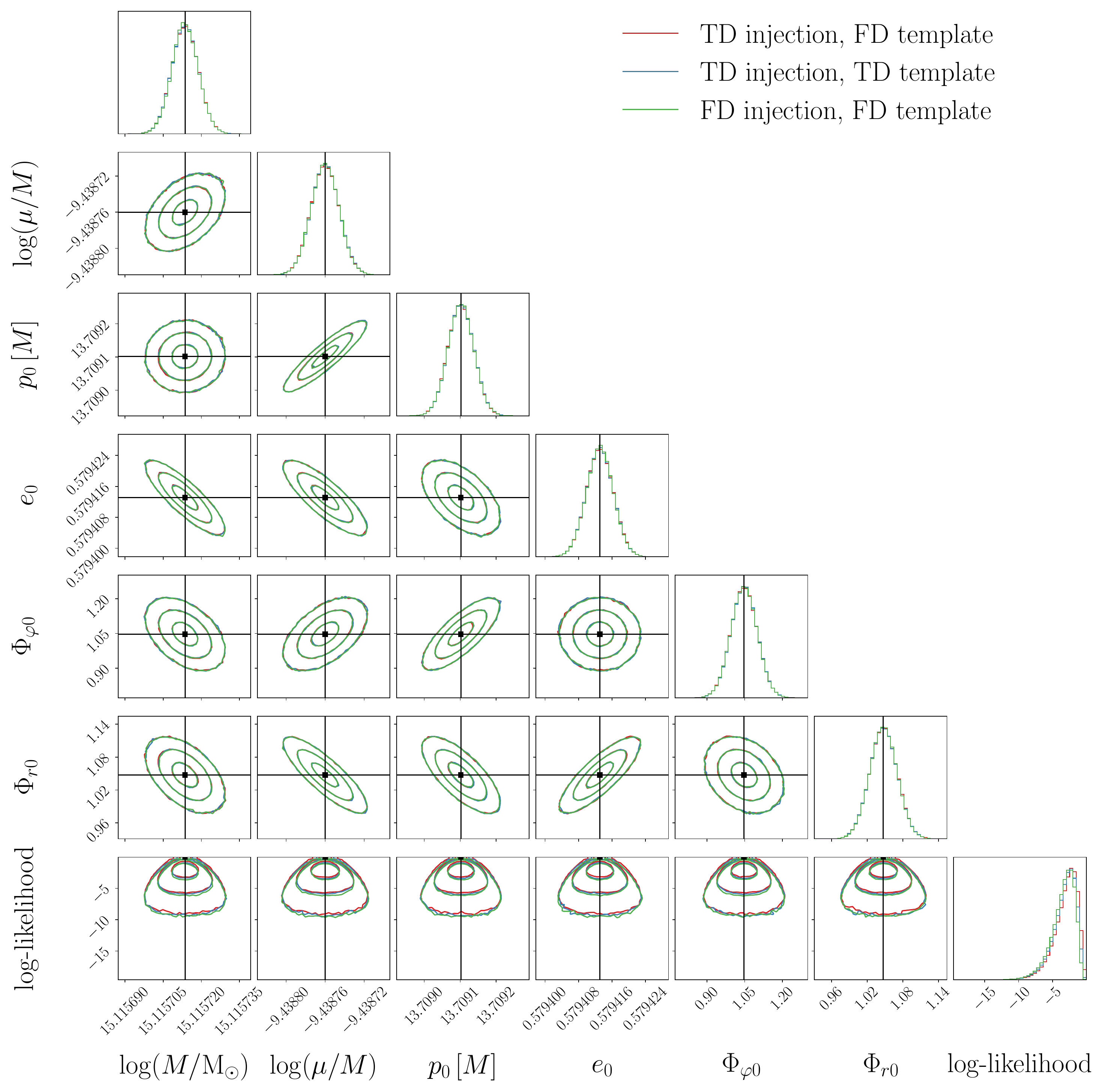}
\label{fig:posterior_td_fd}
\end{figure}

We use Bayesian parameter estimation to assess whether the FD waveforms are affected by systematic errors. We use the Eryn package \citep{Karnesis:2023ras} to perform a Markov chain Monte Carlo analysis to estimate the posterior distribution of the EMRI intrinsic parameters.
Since this analysis is computationally expensive and cannot be performed over the 5000 realizations considered in the mismatch analysis, we decided to focus on the EMRI parameters that yield the largest $\mathcal{M} \times {\rm SNR}^2$. For the values of $T=4$ yrs, $\Delta t = 5$ s or $\Delta t = 10$ s, $\epsilon=10^{-5}$ (blue solid and dotted orange histograms), we find that the largest value of $\mathcal{M} \times {\rm SNR}^2$ is reached by the system with parameters reported in Table~\ref{tab:source}. 
The posterior distribution of this system is shown in Figure~\ref{fig:posterior_td_fd} for the three scenarios in which the FD and TD are used as injections or templates. The Kullback-Leibler divergences \citep{KLdiv} between each pair of posteriors are of order $10^{-4}$ for all the considered parameters. Therefore, we conclude that we do not find any significant loss of accuracy or precision for the worst point in parameter space.

\subsection{Computational Cost of the Waveform generation}\label{sec:speed}
The computational cost of the time and frequency domain implementations differ only in the final waveform summation and output domain. Here, we compare the waveform speeds without including the DFT cost. We make this choice to highlight the difference in the waveform generation cost and we provide the speed of the DFT for a reference system. The speed of the waveform is evaluated using the NVIDIA A100 GPU and the computer cluster \href{https://hypatia.aei.mpg.de/cgi-bin/hypatia-index.cgi?p=main}{Hypatia}\footnote{Hypatia consists of 262 GIGABYTE compute nodes (dual-socket, sixteen-core, SMT-enabled AMD EPYC (Naples) 7351 (2.40 GHz), of which 14 provide 8 GB RAM per core (4 GB per core otherwise).}, for GPU  and CPU timings reported below. We warn the reader that the computational cost of the waveform can vary depending on the computing resources used.

\begin{figure}[]
\caption{\emph{Upper panel:} Speed-up factor of the GPU waveform generation in the time domain (TD) versus frequency domain (FD) over the parameter space of EMRI systems and for different observation time-spans $T$, sampling interval $\Delta t$ and mode content~$\epsilon$. The speed-up factor is given by the ratio of the TD generation speed divided by the FD generation speed per each parameter realization.
The different parameter realizations are drawn uniformly from the following ranges: $\log (M/{\rm M}_{\odot})\in [\log 10^{5},\log  10^{7})$, $\log (\mu/M) \in [\log 10^{-6},\log 10^{-4})$ , $e_0\in [0.001, 0.7)$, $\Phi_{\varphi 0}, \Phi_{r 0} \in [0, 2\pi)$. \emph{Lower panel:} Speed-up factor as a function of central black hole mass $M$ and initial eccentricities $e_0$ for $T=4$ years, $\Delta t=5$ seconds and mode content~$\epsilon=10^{-5}$.
}
\centering
\includegraphics[width=0.8\linewidth]{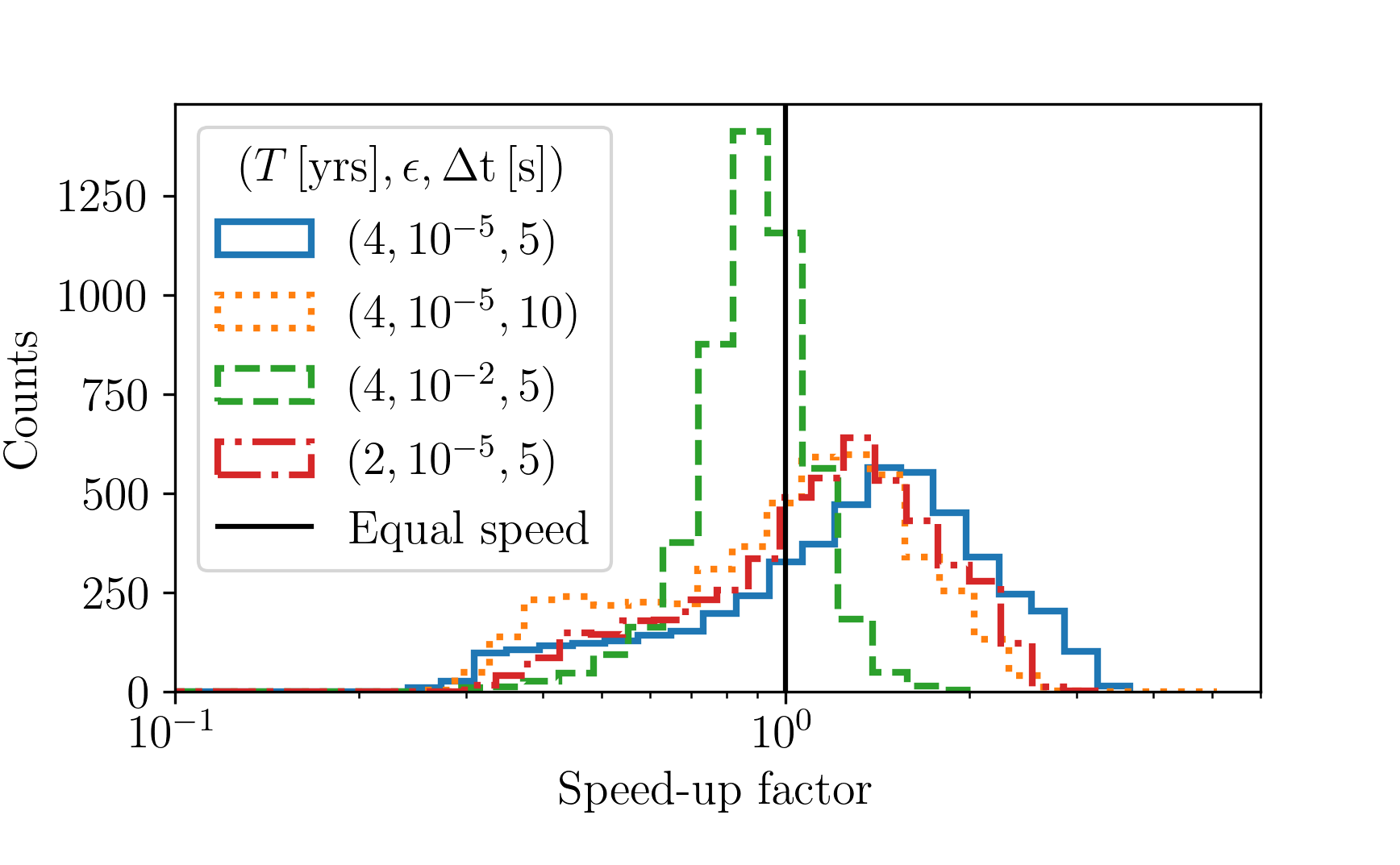}
\includegraphics[width=0.8\linewidth]{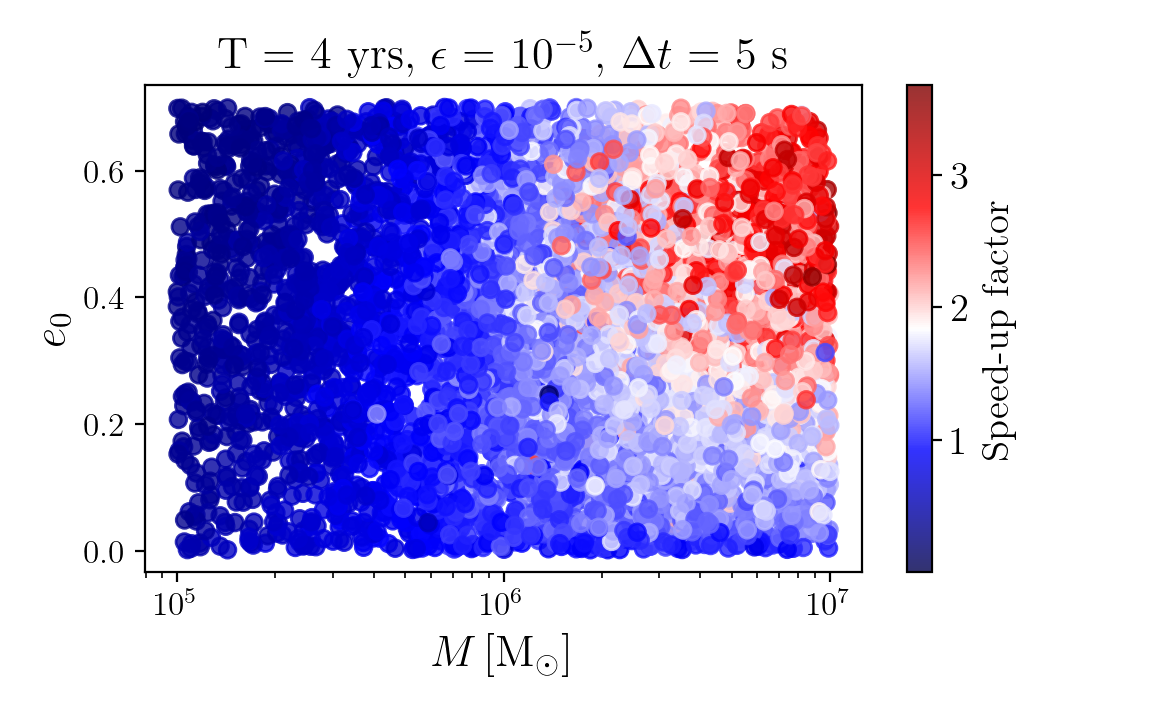}

\label{fig:speedup}
\end{figure}

By drawing 5000 realizations from the parameter space as discussed in the previous section, we show the speed-up factor given by the ratio of the TD and FD waveform evaluation speeds per each realization. The results for the GPU speed-up factor are shown in Figure~\ref{fig:speedup}. The upper panel shows the speed-up as a function of sampling interval, observation time and mode content, while the lower panel shows the dependence on eccentricity and central mass for the fiducial system. Note that the median speeds of all considered configurations are $0.044$ and $0.055$ seconds for the FD and TD waveforms, respectively. The distributions shown in the upper panel of Figure~\ref{fig:speedup} depend on the priors chosen for the parameter space.

Since the mode threshold $\epsilon$ controls the mode content, we can deduce that the FD waveform generation is faster than the TD one when the harmonic content is larger, i.e., $\epsilon$ is smaller (see solid blue and dashed green histograms). This is a key result because future expansion of the {FEW} package will implement EMRI systems with a central rotating black hole where the number of harmonics is expected to increase by an order of magnitude.

As shown by the solid blue and red dashed-dotted histograms in the upper panel of Figure~\ref{fig:speedup}, the FD waveform creation is faster than the TD one for longer signals. This is expected since the TD generation increases with the number of time points. Instead, the FD generation slows down with an increasing {frequency resolution}. This is confirmed by comparing the orange and blue histograms, which differ only by the time sampling interval $\Delta t$. The FD generation for our fiducial case is two times faster for systems with $M>2\times 10^6$ and $e_0>0.2$. 
The performance of the FD waveform compared to the TD one improves with increasing eccentricity and central mass $M$. However, the speed-up trend shown in the lower panel of Figure~\ref{fig:speedup} moves to higher eccentricities and higher masses when $T=2$ years, resulting in a smaller speed-up region. In the limit of short signal durations we expect the TD generation to be faster than the FD one, and contrary in the long duration regime. 

The CPU waveform generation can take up to hundreds of seconds depending on $(T,\epsilon,\Delta t)$. Due to the significantly higher computational cost, we do not present the full scan of the parameter space, but from 10 parameter realizations we conclude that for $(T,\epsilon,\Delta t)=(4 \, \rm{yrs}, 10^{-2}, 10 \, {\rm s})$ 
the FD waveform generation is on average 5 times faster than the TD one on CPUs and takes 5 seconds in median.
For reference, we also report in Table~\ref{tab:timing_source} the main findings after timing the FD and TD speeds for the system of Table~\ref{tab:source} with $(T,\epsilon,\Delta t)=(4 \, \rm{yrs}, 10^{-5}, 10 \, {\rm s})$.
The FD waveform is seven times faster than the TD waveform on CPUs, but only 1.3 times faster on GPUs. The DFT operation is particularly expensive on CPUs taking 7 seconds compared to the 0.4 milliseconds on a GPU. 

\subsection{Frequency downsampling}
One of the main advantages of the FD generation is that we can evaluate the waveform on a sparser frequency array than the one imposed by the frequency resolution $1/T$. We use an evenly-spaced frequency array that spans only the region where the injected waveform is non-zero $f\in[0, f_{\rm max}]$. We change the frequency spacing to downsample the array and we define the total {number of frequency bins} to be $N_f=f_{\rm max}/\Delta f$. We show in Table~\ref{tab:timing_source} 
the computational cost of the waveform generation in frequency and time domain on GPU and CPU obtained for an EMRI system with the parameters specified in Table~\ref{tab:source} and $(T,\epsilon,\Delta t)=(4 \, \rm{yrs}, 10^{-5}, 10 \, {\rm s})$.
As shown in Table~\ref{tab:timing_source} the frequency downsampling strongly affects the CPU speed-up factors, reducing the cost by a factor of 34. For GPUs, instead, the frequency downsampling mildly affects the speed-up factor between FD and TD generation since the GPU timing is already of the order of tens of milliseconds.

\begin{table}[]
\caption{Computational cost of the waveform generation in frequency and time domain on GPU and CPU obtained from an EMRI system with parameters given in Table~\ref{tab:source} and $(T,\epsilon,\Delta t)=(4 \, \rm{yrs}, 10^{-5}, 10 \, {\rm s})$. The speed of the DFT and of the downsampled frequency domain generation is also reported. The last row indicates the size of the frequency array.}
\vspace{0.5cm}
\centering
\begin{tabular}{rccc}
Operation = & FD waveform (Downsampled)& TD waveform & DFT 
\\[0.2cm]
 Speed CPU [s] = & $13.7 \, \, (0.4)$ & $90.6$ & $7.1$  \\[0.2cm]
 Speed GPU [s] = & $0.048 \, \, (0.035)$ & $0.064$ & $0.0004$ \\[0.2cm]
 Number of frequency bins $N_f$ = & $6311631 \, \,(7864)$ & $6311631$ & $6311631$ \\
\end{tabular}

\label{tab:timing_source}
\end{table}

One of the downsides of reducing the {frequency resolution} is the loss of accuracy in the inner product calculation, with a possible impact on parameter inference. Here, we study the loss of accuracy as a function of the {number of frequency bins}. As a proof of concept, we consider the source with the parameters reported in Table~\ref{tab:source} and obtain the posterior distributions using the FD waveform with different input frequencies and with $(T,\epsilon,\Delta t)=(4 \, \rm{yrs}, 10^{-2}, 10 \, {\rm s})$. In this analysis, we did not include windowing since the injected and template waveforms are in the same domain. The absence of windowing also affects the total SNR, so we rescaled the distance to keep the SNR reported in Table~\ref{tab:source} for the finest frequency resolution.
The input array is given by evenly-spaced frequencies between zero and a maximum frequency $f_{\rm max}$. The spacing is set by choosing the total {number of frequency bins} $N_f$. 

\begin{figure}[]
\caption{Posterior distributions for an EMRI system with parameters given in Table~\ref{tab:source}. The posteriors are obtained using the frequency domain waveform with a specified frequency array $f\in[0,f_{\rm max}]$ with $N_f$ {number of frequency bins}. The solid blue posterior considers a maximum frequency set by $f_{\rm max}=1/(2\Delta t)$, whereas the maximum frequency of the dashed orange and dash-dotted green posteriors is set by the maximum frequency present in the signal. The computational cost of the likelihood evaluation is reported in seconds for GPU and CPU. The mode content is fixed to $\epsilon= 10^{-2}$, and the injected waveforms are shown in Figure~\ref{fig:spectrum_downsampled}.
}
\centering
\includegraphics[width=0.95\linewidth]{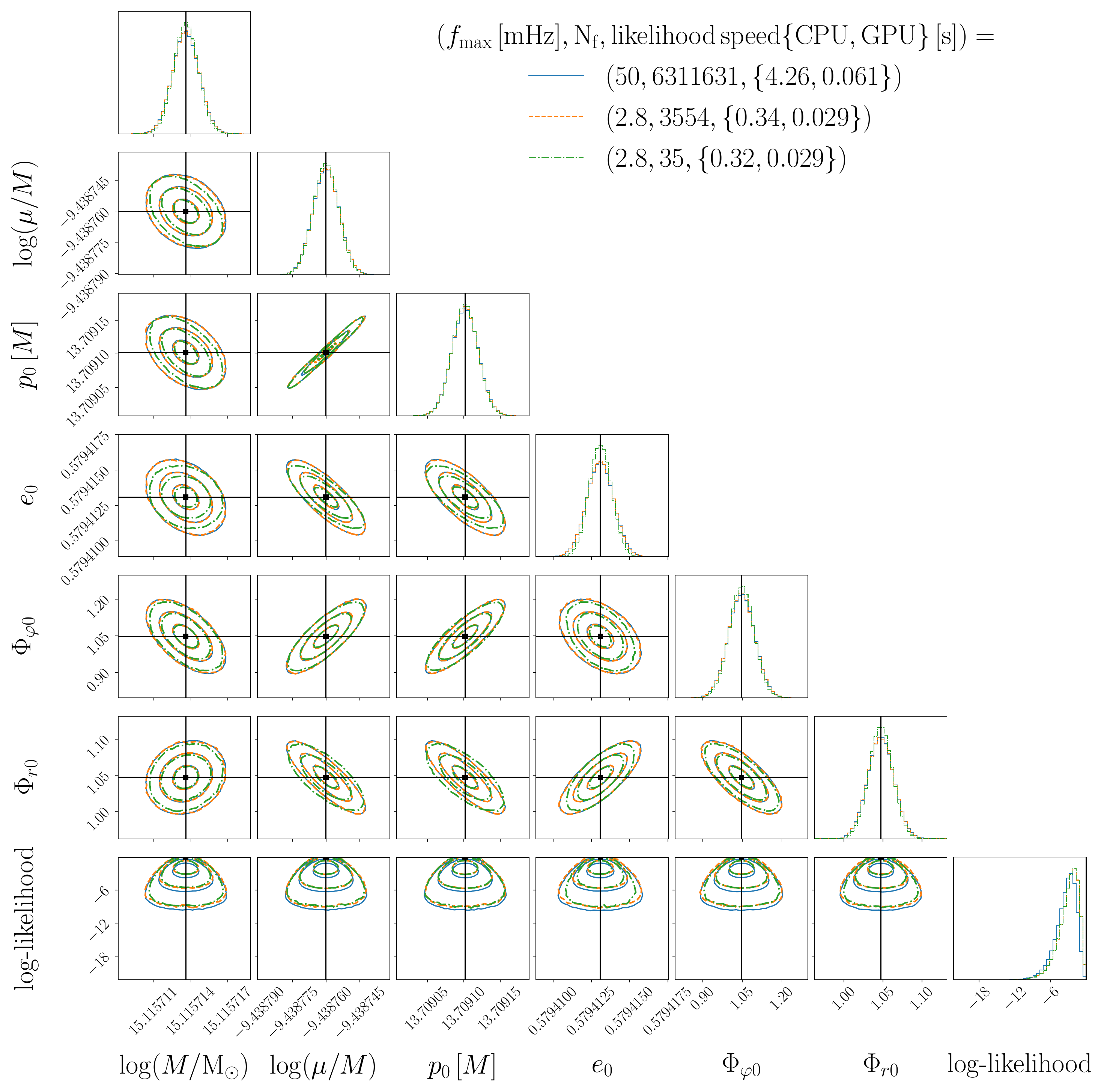}

\label{fig:posterior_downsample}
\end{figure}

In Figure~\ref{fig:posterior_downsample}, we show the posteriors obtained with different levels of downsampling and report the CPU and GPU likelihood speeds. The solid blue posterior is obtained using a frequency array defined by the DFT resolution {$f_{\rm max}=1/(2\Delta t)$} and $N_f=6311631$, whereas the maximum frequency of the dashed orange and dash-dotted green posteriors is set by the maximum harmonic frequency present in the signal $f_{\rm max}=2.8$ mHz. The only difference between the three cases is due to the approximation of the innner product in calculating the likelihood. 
Even using only 0.05\% of the frequencies, the downsampled analysis with $N_f=3554$ (dashed orange histograms) provides a posterior indistinguishable from the one of the full frequency array (solid blue). When further decreasing $N_f$ to 35 the downsampling starts to affect the obtained posteriors. 
The frequency downsampling leads to a faster likelihood evaluation which drops from 4.26 seconds for the full array to 0.34 for CPUs. Further reducing the {number of frequency bins} does not significantly improve the speed, highlighting that the bottleneck of the likelihood computation might be due to other waveform computation steps, such as mode selection \citep{Katz:2021yft}. 
The GPU likelihood timing is reduced by a factor of two when using downsampling.

\begin{figure}[]
\caption{Characteristic strain spectrum $|f \tilde h (f)|^2$ obtained with the frequency domain waveform evaluated on different frequency arrays with $N_f$ frequencies. The solid blue {spectrum} considers a maximum frequency set by $f_{\rm max}=1/(2\Delta t)=50$ mHz, whereas the maximum frequency of the dashed orange and dash-dotted green spectrum is set by the maximum frequency present in the signal $f_{\rm max}=2.8$ mHz.
The parameters of the EMRI system are reported in Table~\ref{tab:source}, and the posteriors for these different levels of frequency downsampling are shown in Figure \ref{fig:posterior_downsample}.
}
\centering
\includegraphics[width=0.95\linewidth]{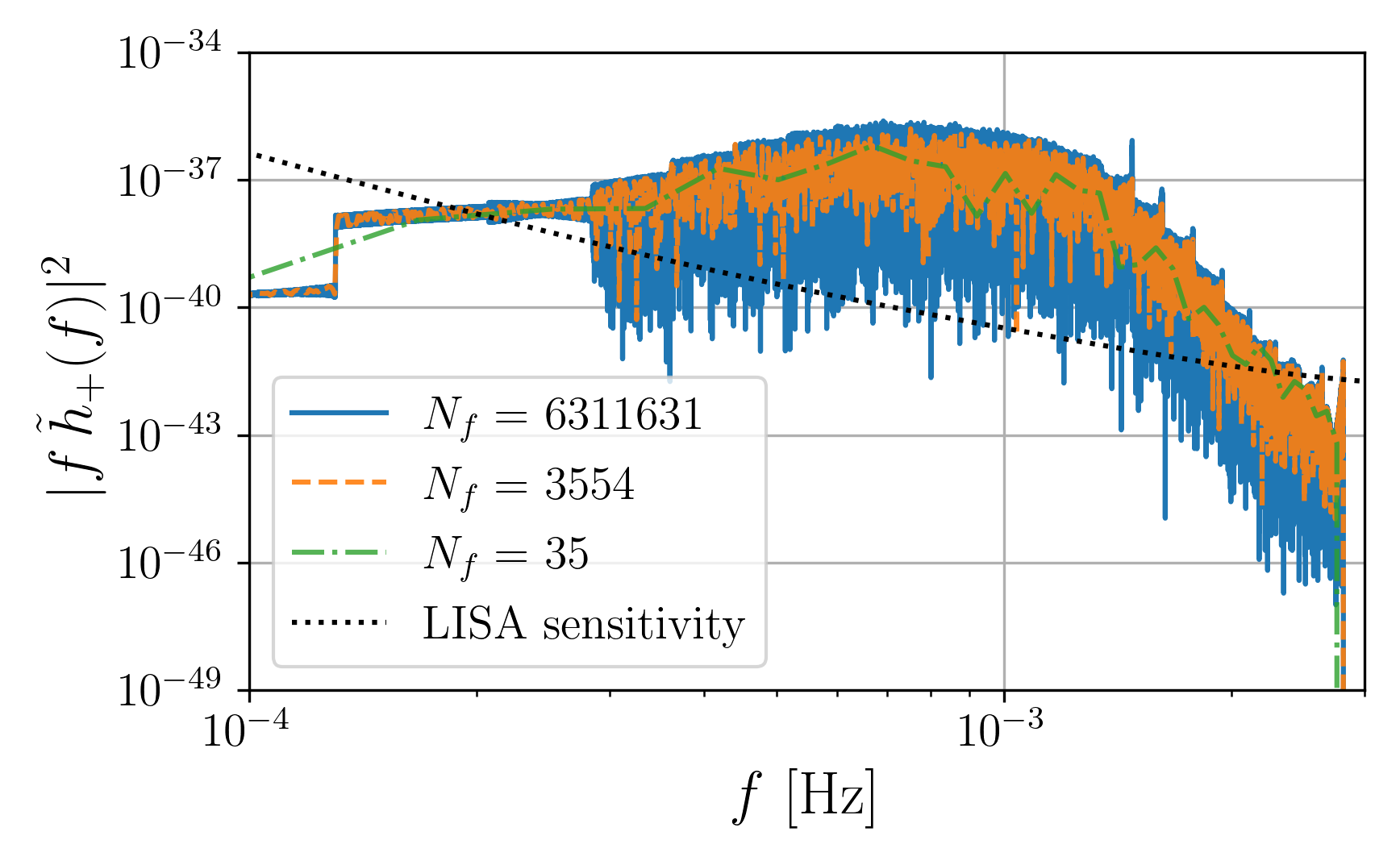}

\label{fig:spectrum_downsampled}
\end{figure}

The reason for the deviation in the posteriors can be explained by looking at the absolute square of the characteristic strain $|f \tilde h (f)|^2$ in Figure~\ref{fig:spectrum_downsampled}. The FD waveform spectrum obtained with $N_f=3554$ (dashed orange line) well represents the spectrum obtained with the full array $N_f=6311631$ (solid blue line) in the frequency range where the signal is above the LISA sensitivity. Instead, the spectrum obtained with $N_f=35$ (dotted green line) has a very sparse frequency array that cannot describe all the features present in the EMRI signal. We conclude that the usage of frequency downsampling can be an important tool for exploratory studies of EMRI parameter inference on CPUs. This motivates the FD implementation as an alternative to the TD one for CPU facilities. However, we stress that it is essential to check the validity of the downsampling before using it.

\section{Discussion and Conclusions}
In this work, we presented a ready-to-use implementation of an EMRI frequency domain waveform, and we compared its accuracy and performance against its time domain counterpart. We found the largest mismatch to be at $3 \times 10^{-3}$, and 95\% of the generated waveforms over the EMRI parameter space have a mismatch below $10^{-4}$. We further investigated the accuracy with a Bayesian analysis of the worst point in the parameter space and found no visible bias in the recovered parameters.

The frequency domain's computational cost on GPUs is lower than the time domain cost for central MBH masses of $>10^{6}$ and for initial eccentricities $>0.2$ and high numbers of harmonic modes. This suggests that the frequency domain model might be more suitable when including a large number of harmonics and we expect the FD implementation to be an important alternative to the TD one when extending the waveform model to Kerr spacetimes, where the number of modes is expected to increase by an order of magnitude. Thanks to the modularity of the FD implementation, the current model can be extended to Kerr spacetimes once amplitude and trajectories are available.
These results also have implications for the data analysis applications of the FD waveform model. If we expect the search phase of EMRIs to be conducted using the fastest model, for instance, with a lower mode content, then the TD generation may be preferable. However, including all the modes might be important when performing tests of general relativity, and therefore the FD waveform could be used in this scenario.

The speed-up factor of the waveform model on CPUs is $\approx 7$ times faster than the time domain version for the considered system, and, in median, $5$ times faster for different configurations. However, the waveform evaluation still takes order of tens of seconds, making the FD waveform not fast enough to perform EMRI parameter inference on CPUs. The frequency domain formulation allows us to choose the input frequency array giving the possibility of downsampling. This was explored for an EMRI system in which we found that the CPU likelihood evaluation can be reduced to $0.3$ seconds when reducing the {number of frequency bins} to be evaluated.
This allows users without GPU resources to run EMRI parameter inference with a fully relativistic waveform for the first time. 
Frequency downsampling can be used for exploratory studies aimed at assessing the uncertainties that might be realized in parameter estimation. However, this must be done carefully especially when injecting noise. The noise must be scaled appropriately to avoid obtaining misleading results, and the possible downsampling will depend on the EMRI parameters. Downsampling will in general not be possible when analyzing real data, but it can be used as a tool for understanding EMRI parameter estimation.

In the current setup, the CPU evaluation time converges to values determined by the mode selection. In future works, we plan to improve this module, therefore enhancing the performance of the CPU FD implementation.
Other approaches that use the sparsity of the frequency array to speed up have already been presented in \citep{Cornish:2021lje,Zackay:2018qdy}. However, the non-monotonic modes and the large number of harmonics pose challenges in applying such techniques to EMRI waveforms. It is not clear yet whether relative binning or heterodyning will significantly speed up the parameter estimation of EMRI sources.

In this work, we did not include the effect of the LISA response function. This is crucial in using the FD waveforms for realistic data analysis studies. A frequency domain response was already presented in \cite{Marsat:2018oam}. However, it still needs to be assessed whether the approximations used in \cite{Marsat:2018oam} apply to EMRIs. This should be investigated in future work.

Looking to the future, we note that post-adiabatic corrections to the waveform phase are important for precision science with EMRIs. These corrections are known for quasi-circular inspirals \citep{Wardell:2021fyy}, and the path to eccentric orbit calculations was laid out in \citep{Leather:2023dzj}. Although the necessary offline second-order self-force calculations are computationally demanding, the online generation of the inspiral is structurally the same, and as fast, as the adiabatic model \citep{Miller:2020bft}. With the modular FD implementation presented in this work the FD post-adiabatic waveform will be available as soon as the post-adiabatic corrections to the inspiral phases are known.

Finally, a full realistic data analysis pipeline has not been developed for EMRIs yet (see, however, \cite{Babak_2009,Cornish_2011}). We stress the importance of developing waveform models in the time, frequency and time-frequency domains in order to explore the advantages and disadvantages of each domain \citep{Cornish2020Wavelet}.

\section*{Conflict of Interest Statement}

The authors declare that the research was conducted in the absence of any commercial or financial relationships that could be construed as a potential conflict of interest.

\section*{Author Contributions}

LS: Conceptualization, Data curation, Formal analysis, Investigation, Methodology, Project administration, Resources, Software, Supervision, Validation, Visualization, Writing – original draft.

MLK: Conceptualization, Data curation, Formal Analysis, Investigation, Methodology, Resources, Software, Supervision, Validation, Writing and Editing. 

AJKC: Conceptualization, Funding acquisition (article processing charges), Methodology, Project administration, Supervision, Writing - review and editing.

SAH: Conceptualization, Formal analysis, Funding acquisition (at least of the portion relevant to developing the waveform data / fluxes), Writing - review and editing.

NW: Conceptualization, Data curation, Formal analysis, Software, Writing - review and editing.

JET: Writing - review and editing.

CEAC-B: Writing - review and editing.

JRG: Writing - review and editing.

\section*{Funding}
AJKC and JET acknowledge support from the NASA LISA Preparatory Science grant 20-LPS20-0005. NW acknowledges support from a Royal Society - Science Foundation Ireland University Research Fellowship.
SAH gratefully acknowledges support from US National Science Foundation Grant PHY-2110384, and past support from NASA Astrophysics Theory Program Grant 80NSSC18K109.
This publication has emanated from research conducted with the financial support of Science Foundation Ireland under Grant numbers 16/RS-URF/3428,
17/RS-URF-RG/3490 and URF/R/221054. 
CEAC-B is supported by STFC studentship 2446638. 

\section*{Acknowledgments}
LS acknowledges Ollie Burke for the very helpful discussions on the impact of windowing in LISA data analysis. AJKC thanks Gabriel Piovano for useful comments on the manuscript. This work makes use of the Black Hole Perturbation Toolkit~\citep{BlackHolePerturbation}.

\section*{Data Availability Statement}
This study is fully reproducible using the publicly available scripts at \url{https://github.com/lorenzsp/EMRI_FrequencyDomainWaveforms}.

\bibliographystyle{Frontiers-Harvard} 
\bibliography{test}

\end{document}